\documentclass[aps,prb,twocolumn,superscriptaddress]{revtex4}
\usepackage{bm}
\usepackage{amsmath}
\usepackage{amssymb}
\usepackage{caption}
\usepackage{graphicx}
\graphicspath{{img_1/}}
\usepackage{color}
\input{epsf}
\usepackage[english]{babel}
\begin{document}
\title{Polaronic effects induced by non-equilibrium vibrons
in a single-molecule transistor}

\author{O. M. Bahrova}
\affiliation{B. Verkin Institute for Low Temperature Physics and
Engineering of the National Academy of Sciences of Ukraine, 47
Prospekt Nauky, Kharkiv 61103, Ukraine}

\author{S. I. Kulinich}
\affiliation{B. Verkin Institute for Low Temperature Physics and
Engineering of the National Academy of Sciences of Ukraine, 47
Prospekt Nauky, Kharkiv 61103, Ukraine}

\author{I. V. Krive}
\affiliation{B. Verkin Institute for Low Temperature Physics and
Engineering of the National Academy of Sciences of Ukraine, 47
Prospekt Nauky, Kharkiv 61103, Ukraine}
\affiliation{Physical Department, V. N. Karazin National
University, Kharkiv 61022, Ukraine}
\affiliation{Center for Theoretical Physics of Complex
Systems, Institute for Basic Science (IBS), Daejeon 34051, Republic
of Korea}

\begin{abstract}
Current-voltage characteristics of a single-electron transistor
with a vibrating quantum dot were calculated assuming vibrons to
be in a coherent (non-equilibrium) state. For a large amplitude of
quantum dot oscillations we predict strong suppression of
conductance and the lifting of polaronic blockade by  bias voltage
in the form of  steps in $I-V$ curves. The height of the steps
differs from the prediction of the Franck-Condon theory (valid for
equilibrated vibrons) and the current saturates at lower voltages
then for the case, when vibrons are in equilibrium state.
\end{abstract}

\maketitle

\section{Introduction}

Tunneling spectroscopy is a well-known method to study of
electron-phonon interaction in bulk metals (see e.g.
Ref.\cite{Yanson1}). Electron transport spectroscopy can be used
for studying of vibration properties of molecules in
single-molecule-based transistors \cite{McEuen,Utko}.
Current-voltage characteristics of single electron transistors
(SETs), where fullerene molecule \cite{McEuen}, suspended
single-wall carbon nanotube \cite{van der Zant, Leturcq, babic} or
carbon nano-peapod \cite{Utko} are used as a base element,
demonstrate at low temperatures additional sharp features (steps)
at bias voltages $eV_n\simeq n\hbar\omega$ ($\omega$ is the
angular frequency of vibrational degree of freedom). The simplest
models (see e.g. review Ref.\cite{krivepalevskij}) that describe
step-like behavior of $I-V$ curves are based, as a rule, on a
theory where phonon excitations are dispersion-less (vibrons with
a single frequency) and they are assumed to be in equilibrium with
the heat bath at temperature $T$ (bulk metallic electrodes can
play the role of this heat bath). Steps in current-voltage
dependencies (equidistant peaks in differential conductance) are
associated with the opening of inelastic channels of electron
tunneling through vibrating quantum dot. For strong
electron-vibron interaction these models predict: (i)
Franck-Condon blockade \cite{von  Oppen} (exponential suppression)
of conductance at low temperatures $T\ll\hbar\omega$ , and (ii)
non-monotonous temperature dependence of conductance. All these
effects were observed in experiments \cite{McEuen, Utko}.

When coupling of vibron subsystem to the heat bath is weak and
vibrons are not in equilibrium during the time of electron
tunneling through the system, their density matrix can not be in
the Gibbs form and it has to be evaluated from the solution of
kinetic equations. This problem can be solved only numerically
(see e.g.\cite{kinaret}). There are only few papers \cite{mitra,
kit, kit2}, where vibrons in electron transport in SET were
considered as non-equilibrated. In Ref.\cite{kit} it was assumed
that vibron subsystem is in a coherent state. In the approach used
in the cited paper, the density matrix of coherent state was
time-independent, that contradicts Liuville-von Neumann equation
for density matrix of {\it noninteracting} vibrons. Therefore the
results of this approach are questionable and the problem of
electron transport through a vibrating quantum dot with coherent
vibrons has to be re-examined.

In our paper we consider single-electron transistor with vibrating
quantum dot, where vibronic subsystem is described by time-
dependent density matrix. Physically this approach corresponds to
coherent oscillations of quantum dot treated as harmonic quantum
oscillator. Coherent states of harmonic oscillators are well known
in physics (see e.g. Refs\cite{coh}). In tunnel electron transport
they are appeared, for instance, in weak superconductivity
(Josephson current through a vibrating quantum dot, see
Ref.\cite{Zazunov} and referencies therein). Last years coherent
states of photons ("Schroedinger cat" states) coupled to qubits
and qubits formed by the coherent photon states became a hot topic
of studies in quantum computing science (see e.g. review
Ref.\cite{Girvin}).

\begin{figure}
\includegraphics[width=0.85\columnwidth]{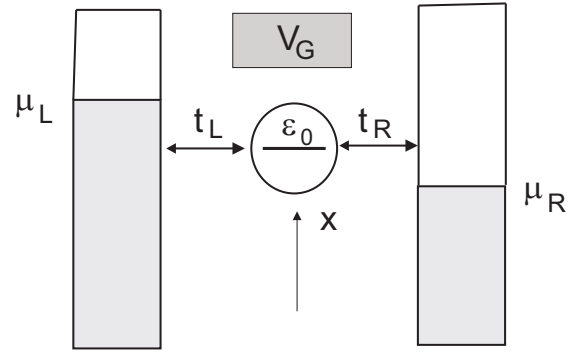}
\caption{Sketch of the single-electron transistor. A vibrating
one-level ($\varepsilon_0$ is the level energy) quantum dot
(macromolecule) is placed between two bulk electrodes biased by
the voltage $V$. The dot tunnel couples ($t_L=t_R=t_0$ is the
tunneling amplitude) to the leads with the chemical potentials
$\mu_{L,R},\;\mu_L-\mu_R=eV$ and the temperature $T$. The gate
voltage $V_G$ is set $\varepsilon_0(V_G)=\varepsilon_F$, where
$\varepsilon_F$ is the Fermi energy, to get maximal current. The
dot oscillates in $x$ direction perpendicular to the electric
current flow. QD oscillations are modelled  by the coherent state
of one dimensional harmonic oscillator.} \label{model}
\end{figure}

The model device we are interesting in is depicted in Fig.1. It
consist of two bulk electrodes, source (Left) and drain (Right)
leads, with chemical potential biased by voltage $\mu_L-\mu_R=eV$
and a single level quantum dot (QD), which oscillates in the
direction ("$x$") perpendicular to the direction of electron
current flow. Gate voltage, $V_G$, is adjusted to maximum tunnel
current $\varepsilon_0(V_G)=\varepsilon_F$, where
$\varepsilon_0(V_G)$ is the dot level energy and $\varepsilon_F$
is the Fermi energy of the leads. For simplicity we consider
tunneling of spinless electrons in a symmetric junction and it is
assumed that the vibration of QD does not change tunneling matrix
elements $t_L=t_R=t_0$ . In our paper we consider the process of
sequential electron tunneling, when $\max(eV, T)\gg\Gamma$, where
$\Gamma\propto |t_0|^2$ is the level width (characteristic energy
of tunnel coupling dot-leads). Our model device can simulate , for
instance, SET based on a suspended single-wall carbon nanotube.

We use density matrix approach to calculate periodic in time
current through the device (the period $T_0= 2\pi/\omega$ is
determined by the angular frequency $\omega$ of QD oscillations).
In order to calculate current-voltage dependencies, we numerically
average the current over $T_0$. It is shown that the
zeroth-harmonic (time-independent) contribution dominates in the
Fourier series for the current. Therefore a simple analytic
equation for dc electric current (analogous to the current through
vibrating QD with equilibrated vibrons) is presented. This formula
agrees with our numerical calculation with a high accuracy.

We show that $I-V$ characteristics of a single-electron transistor
with coherent vibrons are a step-like function of bias voltage and
they do not depend on the phase of coherent state parameter. At
large amplitudes of dot oscillations the conductance is strongly
suppressed ("polaronic blockade") regardless the strength of
electron-vibron interaction. The heights of the steps and the
characteristic voltage of current saturation strongly differ from
the prediction the Franck-Condon theory. In particularly the
lifting of polaronic blockade occurs at lower voltages than the
lifting of Franck-Condon blockade.

\section{Hamiltonian and equation for density matrix}
The Hamiltonian of the system (see schematic picture of our device, Fig.1)
consists of four terms,
\begin{equation}\label{1}
H= H_l+ H_{dot}+H_{v-d}+ H_{tun},
\end{equation}
where $H_l, H_{dot}$ are the Hamiltonians of the non-interacting
electrons in the leads and the dot correspondingly,
\begin{equation}\label{2}
H_l=\sum_{k,\kappa}\varepsilon_{k,\kappa}
a^\dag_{k,\kappa}a_{k,\kappa},\, H_{dot}=\varepsilon_0 c^\dag c,
\end{equation}
$a_{k,\kappa}^\dag (a_{k,\kappa})$ is the creation (annihilation)
operator (with standard anti-commutation relations) of electron in
the lead $\kappa=L,R$ with momentum $k$ and energy
$\varepsilon_{k,\kappa}$ , $c^\dag(c)$ is the creation
(annihilation) operator of electron state in the dot with the
energy $\varepsilon_0$.

Hamiltonian $H_{v-d}$ describes the vibronic subsystem and the
interaction between electrons and vibrons,
\begin{equation}\label{4}
\hat H_{v-d}=\frac{p^2}{2m}+\frac{m\omega^2 x^2}{2}+\Delta
xc^\dag c.
\end{equation}
In Eq.(\ref{4}) $x,p$ are the canonical conjugating operators of
coordinate and momentum, $[x,p]=\imath \hbar,\, \omega, m$ are
the frequency of dot oscillations and the mass of the dot,
$\Delta$ is the electron-vibron coupling constant.

The Hamiltonian $H_{tun}$ describes the tunnelling of electrons
between the dot and the leads and it takes the standard
form,
\begin{equation}\label{5}
 H_{tun}=\sum _{k,\kappa} t_\kappa a^\dag_{k,\kappa}c+\text{H.c.},
\end{equation}
where $t_\kappa$ is the tunnelling amplitude. In what follows we
restrict ourselves to the symmetric case, $t_L=t_R=t_0$.

It is convenient to perform the unitary transformation,
$UHU^\dag\rightarrow H$, with $U=\exp[i\lambda p c^\dag c]$ and $
\lambda=\Delta/\hbar m \omega^2$. After this transformation the
dot-vibron Hamiltonian $H_{v-d}$ (Eq. (\ref{4})) takes the
diagonal form,
\begin{equation}\label{7}
H_{v-d}\rightarrow H_v=\frac{p^2}{2m}+\frac{m\omega^2 x^2}{2},
\end{equation}
while the tunnelling Hamiltonian, $H_{tun}$, is
transformed to the equation,
\begin{equation} \label{8}
H_{tun}\rightarrow H_{tun}=t_0\sum_{k,\kappa}\text e^{-\imath\lambda p}
a^\dag_{k,\kappa}c+\text{H.c.}.
\end{equation}

The quantum consideration of electron-vibron interacting system is
based in what follows on the approximation that the density matrix
of the system is factorized to direct product of the leads
equilibrium density matrix, the vibron  density matrix and the
density matrix of the dot,
\begin{equation}\label{9}
\rho \approx\rho_l\otimes\rho_v\otimes\rho_{dot}.
\end{equation}
This approximation corresponds to the case of sequential electron
tunneling, which holds when $\text{max}\{eV,T\}\gg \Gamma$, where
$\Gamma$ is the electron level width, $T$ is the temperature and
$V$ is the biased voltage. In contrast to the previous works (see
e.g. Refs.\cite{krivepalevskij, Shkop}) we will consider
non-equilibrated vibrons. Here we assume that they are described
by a time-dependent coherent state $\vert z(t) \rangle$. Note,
that in Ref.\cite{kit} current-voltage characteristics of a
single-electron transistor were calculated for time-independent
coherent state of vibrons. This assupmtion contradicts to equation
of motion of noninteracting vibrons in our model, where $\vert
z(t)\rangle=\exp\left(-\imath H_{v}t\right)\vert z\rangle,
(\hbar=1)$. Here $\vert z \rangle$ is the eigenfunction of vibron
annihilation operator $b, b\vert z\rangle =z \vert z \rangle$ ($z$
is the complex number). The corresponding density matrix takes the
standard form
\begin{equation}\label{93}
\rho_v(t)=\vert z(t)\rangle \langle z(t)\vert.
\end{equation}
The Liouville-von Neumann equation for the density matrix
\begin{equation}\label{10}
\frac{\partial \rho}{\partial t}+\imath[H_0+H_{tun},\rho]=0,
\end{equation}
where $H_0=H_l+H_v+H_{dot}$, has the formal solution,
\begin{equation}\label{11}
\rho(t)=\rho(-\infty)-\imath\int_{-\infty}^t dt'\text e^{-\imath
H_0(t-t')}[H_{tun},\rho(t')]\text e^{\imath H_0(t-t')}.
\end{equation}
After substitution of Eqs.(\ref{9}), (\ref{11}) into
Eqn.(\ref{10}) and tracing out both the electronic degrees of
freedom of the leads and vibronic degrees of freedom of the dot
one gets
\begin{eqnarray}\label{12}
&&\frac{\partial \rho_{dot}}{\partial
t}+\imath[H_{dot},\rho_{dot}]=\\
&& -\text{Tr}\int_{-\infty}^t dt'[H_{tun}, \text e^{-\imath
H_0(t-t')}[H_{tun},\rho(t')]\text e^{\imath H_0(t-t')}].\nonumber
\end{eqnarray}
Now we can explicitly calculate averages of electronic and
vibronic operators  in our approximation of the factorized density
matrix Eq.(\ref{9}). For equilibrium density matrix of electrons
in the leads we use the standard expression
\begin{equation}\label{13}
\langle a^\dag_{k,\kappa}a_{k',\kappa'}\rangle=
f_\kappa(\varepsilon_{k,\kappa})
\delta_{k,k'}\delta_{\kappa,\kappa'},
\end{equation}
where
$f_\kappa(\varepsilon)=(\exp((\varepsilon-\mu_\kappa)/T)+1)^{-1}$
is the Fermi-Dirac distribution function, $\mu_{L,R}= \mu_0\pm
(eV/2)$ is the electrochemical potential in the lead $\kappa$. The
evaluation of vibronic correlation function $
F(t,t_1;\lambda)=\langle\exp[-\imath \lambda p(t)] \exp[\imath
\lambda p(t_1)]\rangle $ in coherent state representation results
in the equation
\begin{eqnarray}\label{130}
&& F(t,t_1;\lambda)=\text{Tr[ e}^{-\imath \lambda p(t)} \vert
z\rangle\langle z\vert
\text e^{\imath \lambda p(t_1)}]= \nonumber \\
&&\exp\left\{-\lambda^2\left[1-\text e^{\imath
\omega(t-t_1)}\right]-\nonumber\right.\\&& \left.-\lambda z
\left[\text e^{-\imath \omega t}- \text e^{-\imath \omega
t_1}\right]+\lambda z^\ast\left[\text e^{\imath \omega t}- \text
e^{\imath \omega t_1}\right]\right\},
\end{eqnarray}
(in Eq.(\ref{130}) we introduced the dimensionless constant of
electron-vibron interaction, $\lambda\hbar\sqrt{2}/ l_0\rightarrow
\lambda, l_0=\sqrt{\hbar/m\omega}$ is the amplitude of zero-point
oscillations). Parameter $\lambda$ characterises the "degree of
quantumness" of the mechanical subsystem. It can be rewritten in
the form $\lambda=\sqrt{2}l/l_0$, where $l=\Delta/m\omega^2$ is
the characteristic displacement length of classical oscillator.

With the help of Eqs.(\ref{13}), (\ref{130}) Eq.(\ref{12}) can be
represented as follows
\begin{widetext}
\begin{eqnarray}\label{15}
&&\frac{\partial \rho_{dot}}{\partial
t}+\imath[H_{dot},\rho_{dot}]=\frac{\Gamma}{4\pi}\sum_\kappa \int
d\tau\int d\varepsilon  \times\nonumber\\
&&\left\{F(t,t-\tau;\lambda)\text e^{\imath \varepsilon \tau}
\left[1-f_\kappa (\varepsilon)\right]c \text e^{-\imath
H_{dot}\tau}\rho_{dot}(t-\tau)c^\dag \text e^{\imath
H_{dot}\tau}+\right.
\nonumber\\
&&+F(t,t-\tau;-\lambda)\text e^{-\imath \varepsilon \tau} f_\kappa
(\varepsilon)c^\dag \text e^{-\imath H_{dot}\tau}\rho_{dot}(t-\tau)c
\text e^{\imath H_{dot}\tau}-\nonumber\\
&&-F^\ast(t,t-\tau;-\lambda)\text e^{\imath \varepsilon
\tau}f_\kappa(\varepsilon) c \text e^{-\imath
H_{dot}\tau}c^\dag\rho_{dot}(t-\tau)\text e^{\imath
H_{dot}\tau}-\nonumber\\
&&\left.-F^\ast(t,t-\tau;\lambda)\text e^{-\imath \varepsilon
\tau}[1-f_\kappa(\varepsilon)]c^\dag \text e^{-\imath H_{dot}\tau} c
\rho_{dot}(t-\tau)\text e^{\imath H_{dot}\tau}+\text{H.c.}\right\},
\end{eqnarray}
\end{widetext}
where $\Gamma=2\pi\nu t_0^2$ is the level width of electron state
in the dot, $\nu$ is the density of states of the leads, which we
assume to be energy independent (wide-band approximation, see e.g.
Ref.\cite{wingreen}). We notice here that unlike the case of
equilibrated vibrons (see e.g. Ref.\cite{Shkop}), the vibron
correlation function, Eq.(\ref{130}), depends on two times
independently. This means that time-invariance in our system is
explicitly broken. The vibrons in coherent state $\vert z(t)
\rangle$, (which physically describes oscillations of quantum
pendulum) violates time-invariance.

The density operator $\rho_{dot}$ acts in Fock space, which in our
case is a two dimensional space of a spinless electron level in
the dot. The matrix elements of the density operator are
$\rho_0(t)=\langle 0\vert\rho_{dot}(t)\vert
0\rangle,\rho_1(t)=1-\rho_0(t)=\langle 1\vert\rho_{dot}(t)\vert
1\rangle$, where $\vert 1\rangle=c^\dag \vert 0\rangle$ and $\vert
0\rangle$ is a vacuum state. From Eq.(\ref{15}) it follows that
the probability $\rho_0(t)$ satisfies the equation,
\begin{eqnarray}\label{16}
&&\frac{\partial \rho_0}{\partial
t}=\frac{\Gamma}{4\pi}\sum_\kappa\int d\tau \int
d\varepsilon\nonumber\\&& \left\{ F(t,t-\tau;\lambda)\text
e^{\imath(\varepsilon-\varepsilon_0)\tau}
\left[1-f_\kappa(\varepsilon))\right]\left[1-\rho_0(t-\tau)
\right]-\right.\nonumber\\
&&\left.-F^\ast(t,t-\tau;-\lambda)\text
e^{\imath(\varepsilon-\varepsilon_0)\tau}
f_\kappa(\varepsilon))\rho_0(t-\tau)\right\}.
\end{eqnarray}

This equation is strongly simplified after integration over
$\varepsilon$. This integration can be done by using the equation,
\begin{equation}\label{17}
\int d\varepsilon \text e^{-\imath \varepsilon \tau} f_\kappa
(\varepsilon)=-\imath \pi \delta (\tau) +
\text{p.v.}\frac{\imath\pi T\text e^{-\imath \mu_\kappa
\tau}}{\sinh \pi T\tau}.
\end{equation}
In the limit $T\gg\Gamma$ one can neglect the retardation effects
and Eq.(\ref{16}) takes a simple local form,
\begin{equation}\label{18}
-\frac{\partial \rho_0}{\partial t}=M_1(t)\rho_0-M_2(t),
\end{equation}
where
\begin{equation}\label{19}
M_i(t)=1-\frac{1}{2}\sum_n A_n^{(i)}(t)[f_L(\varepsilon_0-n\omega)+
f_R(\varepsilon_0-n\omega)],
\end{equation}
The coefficients $A_n^{(i)}(t)$ are periodic
functions of time (with the period $2\pi/\omega$) and they can
be presented as the Fourier series
\begin{eqnarray}\label{191}
&&\hspace{1cm}A_n^{(i)}(t)=\sum_p a_{n,p}^{(i)}e^{\imath \omega p t},\\
&&a_{n,p}^{(1)}=\frac{1}{\pi}\int_{-\pi}^\pi d\vartheta \text
e^{-\lambda^2(1-\cos \vartheta)}\sin\left( n\vartheta-\frac{\pi
p}{2}\right)\times\\&& \times \sin\left(\lambda^2
\sin\vartheta\right)\cos\left(\frac{p\vartheta}{2}\right)J_p\left(4\lambda
\vert z\vert \sin\frac{\vartheta}{2}\right),\label{192}\nonumber\\
&&a_{n,p}^{(2)}=\frac{1}{2\pi}\int_{-\pi }^\pi d\vartheta \text
e^{-\lambda^2(1-\cos
\vartheta)}\cos\left(\frac{p\vartheta}{2}\right)\times\nonumber\\&&
\times\cos\left(\frac{\pi
p}{2}-n\vartheta+\lambda^2\sin\vartheta\right) J_p\left(4\lambda
\vert z\vert \sin\frac{\vartheta}{2}\right).\label{193}
\end{eqnarray}
In Eqs.(\ref{192}), (\ref{193}) $J_p(x)$ is the Bessel function of
the first kind and we parameterized the coherent state eigenvalue
$z$ in the form $z=\vert z\vert \exp(\imath\varphi)$. Notice, that
that the parameter $\vert z\vert$ determine the amplitude of dot
oscillation.

In the asymptotic ($t\gg 1/\Gamma$) steady state regime of
oscillations the probability $\rho_0(t)$ is a periodic function of
time, $\rho_0(t+T_0)=\rho_0(t)$, and therefore it can be presented
as the Fourier series,
\begin{equation}\label{24}
\rho_0(t)=\sum_n\rho_n e^{\imath\omega n t},\,\rho_{-n}=\rho_n^\ast.
\end{equation}
Then the equation for the Fourier harmonics takes the form
\begin{eqnarray}\label{26}
&&\imath p \rho_{p}=\delta_{p,0}-\rho_p-\frac{1}{2}\sum_n
\left[a_{n,p}^{(2)}-\sum_{k}a_{n,p+k}^{(1)}\rho_{k}\right]\times\nonumber\\
&&\hspace{1cm}\left[f_L(\varepsilon_0-n\omega)+f_R(\varepsilon_0-n\omega)\right].
\end{eqnarray}

We are interested in $I-V$ characteristics of our single-electron
transistor. Therefore we have to calculate time-averaged current
through the system
\begin{equation}\label{20}
I=\frac{1}{T_0}\int_{T_0}^{}J(t)dt,
\end{equation}
where $J(t)=(J_L+J_R)/2$ and the left (L)and right (R) currents in
the system are defined by a standard equation,
\begin{equation}\label{21}
J_\kappa= \eta_\kappa e \text{Tr}\left(\rho\frac{\partial
N_\kappa}{\partial t}\right), \; \; \; N_\kappa=\sum_{k}a_{k,\kappa}^\dag
a_{k,\kappa},
\end{equation}
where $\eta_{L/R}=\pm 1$. With the help of Eq.(\ref{11}) the
expression for the current can be presented in the following form,
\begin{eqnarray}\label{22}
&&J_\kappa=\eta_\kappa\text{Tr}\int_{-\infty}^tdt'\text e^{\imath
H_0(t-t')}I_\kappa \text e^{-\imath H_0(t-t')}[H_{tun},
\rho]+\text{c.c.}, \nonumber\\&&I_\kappa= et_0 \text
e^{-\imath\lambda p}\sum_k ca_{k,\kappa}^\dag.
\end{eqnarray}
The straightforward calculation of Eq.(\ref{22}) yields the
following equation analogous to Eq.(\ref{18})
\begin{equation}\label{23}
\frac{J(t)}{I_0}=-\rho_0(t)P_1(t)+P_2(t),
\end{equation}
where $I_0= e\Gamma/2$ is the saturation current through
a single-level symmetric junction, and
\begin{equation}\label{23a}
P_i(t)=\sum_n A_n^{(i)}(t)\left[f_L(\varepsilon_0-n\omega)-
f_R(\varepsilon_0-n\omega)\right]
\end{equation}
(coefficients $A_n^{(i)}$ are defined in
Eqs.(\ref{191})-(\ref{193})). As it follows from
Eqs.(\ref{24}),(\ref{20}),(\ref{23}), the desired expression for
the average current takes the form
\begin{equation}\label{25}
I=I_0\sum_{n,k}\left[a^{(2)}_{n,k}\delta_{k,0}-a^{(1)}_{n,k}
\rho_k\right]\left[
f_L(\varepsilon_0-n\omega)-f_R(\varepsilon_0-n\omega)\right].
\end{equation}
Notice, that the average current does not depend
on the phase $\varphi$ of coherent state.

\section{Numerical results and discussion}

The results of numerical calculations are presented in Figs.2,3.
As one can see, the plots for coherent vibrons (black dotted
curves) demonstrate step-like behavior of current versus bias
voltage at low temperatures $T\ll\hbar\omega$. This behavior is
similar (however, in general case not identical) to Franck-Condon
steps in $I-V$ curves known for equilibrated vibrons (see e.g.
review paper Ref.\cite{krivepalevskij} and references therein).
The plots for equilibrated and coherent vibrons coincide (see
Fig.2) when the amplitude of oscillations of QD is less or of the
order of the amplitude of zero-point oscillations $l_0$ ($|z|\leq
1$ correspondingly).

\begin{figure}
\includegraphics[width=0.85\columnwidth]{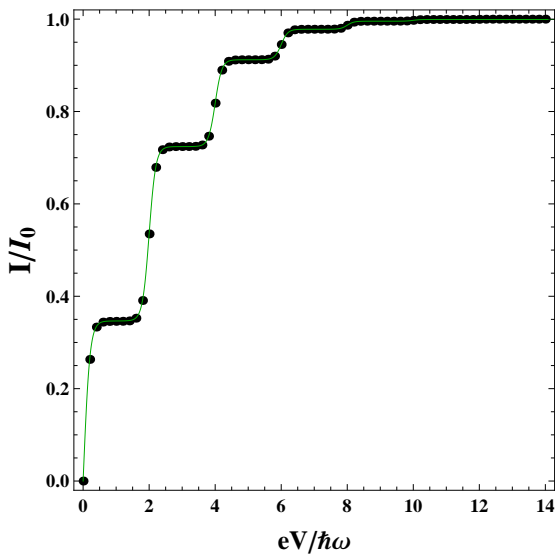}
\caption{The current-voltage dependencies for small value of
coherent state parameter of vibrons, $\vert z\vert=0.25$, and for
strong electron-vibron interaction $\lambda=1$. The black dotted
curve corresponds to numerical calculation of current when the
vibrons are in the coherent state. The thin green curve represents
$I-V$ characteristics when the vibrons are in equilibrium and
characterized by the effective temperature $T^*$ determined by
Eq.(\ref{26}). In calculations the values $T/\hbar \omega =0.05,
\Gamma / \hbar \omega=0.001$ was used. } \label{fig2}
\end{figure}

It is physically clear that in this case both systems are close to
their ground state (the average number of vibrons $\textless
n\textgreater\ll 1$) and there is no difference in the behavior of
coherent and non-coherent vibrons. The strong differences appear
for large amplitudes of oscillations when $|z|\gg 1$ (see Fig.3
where the dotted curve corresponds to vibrons in the coherent
state with parameter $|z|=10$). It is useful to introduce
effective temperature of vibrons  $T^\ast$ by equating the average
number of vibrons in coherent and equilibrium state,
\begin{equation}\label{26}
|z|^2=(\exp(\hbar\omega/T^\ast))-1)^{-1}.
\end{equation}
Then for large amplitudes of oscillations ($|z|\gg 1$) and
moderately strong electron-vibron interaction ($\lambda\sim 1$)
$T^\ast\simeq |z|^2\hbar\omega\gg \lambda^2\hbar\omega$. It is
clear that at these high temperatures of the leads Franck-Condon
steps in $I-V$ characteristics will be smeared out. It means that
coherent vibrons for large amplitudes of QD oscillations lead to
strong suppression of current at low biases and to pronounced
step-like behavior of $I-V$ curves. It is interesting to compare
this behavior with the Franck-Condon theory by assuming that the
vibronic subsystem is hot (it is described by Bose-Einstein
distribution with the temperature $T^\ast$), while the leads are
kept at low temperatures $T\ll\hbar\omega$. The thin curve (green
on-line) in Fig.3 demonstrates this case.
\begin{figure}
\includegraphics[width=0.85\columnwidth]{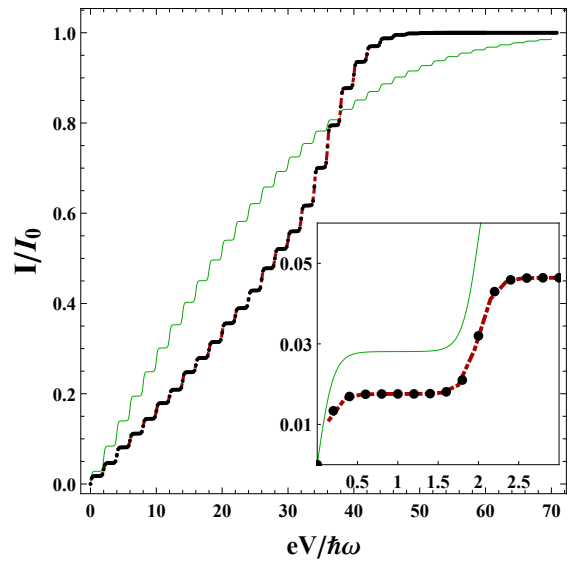}
\caption{$I-V$ plots for the large value of the parameter $\vert
z\vert=10$ . All other parameters are the same as in Fig.2. The
thin green curve corresponds to the case of equilibrated vibrons
with the effective temperature determined by the parameter $\vert
z\vert=10$. The red dash-dotted curve represents calculation of
current in the approximation when $\rho_0=0.5$ (see the text
bellow). Inset shows the region of low voltages.}  \label{fig3}
\end{figure}
We see rather strong differences in current-voltage dependencies:
(i) the height of the steps for coherent vibrons are not regular,
and (ii) the current in the case of coherent vibrons saturates at
lower voltages ($eV_s\simeq |z|\hbar\omega$) than for equilibrated
vibrons.

One can strongly simplify numerical calculations noticing that
coefficient $\rho_0$ (zeroth harmonic) of the Fourier series
Eq.(\ref{24}) in the steady state regime $\rho_0=0.5$  with very
high accuracy, $\sim 10^{-5}$.  Then if we put in Eq. (\ref{25})
$\rho_0=1/2$ and $\rho_p=0$ for $ p\ge 1$, one gets a simple
analytic formula for the average current
\begin{equation}\label{Ias}
I=I_0\sum_{n}a_{n}\left[ f_L(\varepsilon_0-n\omega)
-f_R(\varepsilon_0-n\omega)\right],
\end{equation}
where
\begin{eqnarray}\label{27}
&&a_{n}=\frac{1}{\pi} \int_{0}^{\pi}d\vartheta \text{e}^{-\lambda
^2 (1-\cos{\vartheta})}
\times\nonumber\\&&\hspace{1cm}\times\cos{n\vartheta}
\cos{(\lambda ^2\sin{\vartheta})} J_0\left( 4\lambda \vert z\vert
\sin{\frac{\vartheta}{2}}\right).
\end{eqnarray}
For $\lambda\leq 1$ one can roughly estimate integral
Eq.(\ref{27}) as $ a_n\simeq J_n^2\left( 2\lambda \vert
z\vert\right)$. This allows us to strongly simplify numerical
calculations. Note that Eq.(\ref{Ias}) has the same form as a
well-known equation (see e.g. Ref.\cite{krivepalevskij}) for the
current of spinless electrons through a vibrating QD with
equilibrated vibrons
\begin{equation}\label{Ieq}
I_{eq}=I_0\sum_{n}A_{n}\left[ f_L(\varepsilon_0-n\omega)-f_R(\varepsilon_0-n\omega)\right],
\end{equation}
where now spectral densities $A_n$ are defined by the expression
$\text{Tr [e}^{-\imath \lambda p(t)}\text e^{\imath \lambda
p(0)}\rho_{eq}]=\sum_{n}A_ne^{\imath \omega n t}$.

The dash-dotted curve (red on-line) in the Fig.3 correspond to
calculations by using Eqs.(\ref{Ias}),(\ref{27}). This approximate
calculations coincide with the "exact" numerical calculations with
a high accuracy.

In summary, we have calculated $I-V$ characteristics of a
single-molecule transistor, assuming vibrons of QD (molecule)
oscillations to be in a coherent state. It was shown that $I-V$
curves at low temperatures have a step-like form similar to the
steps that accompany the lifting of Franck-Condon blockade by bias
voltage. However, for large amplitudes of oscillations there are
strong differences in the predictions of the Franck-Condon theory
and our model. By using numerical calculations we found strong
suppression of conductance even for a weak or moderately strong
electron-vibron coupling. The lifting of this coherent
oscillations-induced blockade by a bias voltage occurs at voltages
much lower then the ones predicted by the Franck-Condon theory.

\textbf{\emph{Acknowledgements}}. The authors thank L.Y.Gorelik
and O.A.Ilinskaya for useful discussions. This work is supported
by the National Academy of Sciences of Ukraine (grant No. 4/19-N
and Scientific Program 1.4.10.26.4) and partially by the Institute
for Basic Science in Korea.

\end{document}